\documentclass[12pt,preprint]{aastex}
\usepackage[dvips]{color}

\renewcommand{\d}{\mbox{d}}
\newcommand{\dpp}[2]{\frac{\partial #1}{\partial #2}}
\newcommand{\ddt}[1]{\dpp{#1}{t}}
\newcommand{\ddx}[1]{\dpp{#1}{x}}
\newcommand{\ddy}[1]{\dpp{#1}{y}}
\newcommand{\ddz}[1]{\dpp{#1}{z}}
\newcommand{\DDt}[1]{\frac{\d #1}{\d t}}
\newcommand{\mi}[1]{\mbox{\boldmath$#1$}}

\shorttitle{Simulation of wave field in solar convection zone}
\shortauthors{Parchevsky \& Kosovichev}

\begin{document}
\title{Three dimensional numerical simulations of acoustic wave field in
the upper convection zone of the Sun}

\author{Parchevsky, K.V., and Kosovichev A.,G.}
\affil{455 Via Palou, Stanford University, HEPL, Stanford CA 94305,
USA} \email{kparchevsky@solar.stanford.edu}

\begin{abstract}
Results of numerical 3D simulations of propagation of acoustic waves
inside the Sun are presented. A linear 3D code which utilizes
realistic OPAL equation of state was developed by authors. Modified
convectively stable standard solar model with smoothly joined
chromosphere was used as a background model. High order dispersion
relation preserving numerical scheme was used to calculate spatial
derivatives. The top non-reflecting boundary condition established
in the chromosphere absorbs waves with frequencies greater than the
acoustic cut-off frequency which pass to the chromosphere,
simulating a realistic situation. The acoustic power spectra
obtained from the wave field generated by sources randomly
distributed below the photosphere are in good agreement with
observations. The influence of the height of the top boundary on
results of simulation was studied. It was shown that the energy
leakage through the acoustic potential barrier damps all modes
uniformly and does not change the shape of the acoustic spectrum. So
the height of the top boundary can be used for controlling a damping
rate without distortion of the acoustic spectrum. The developed
simulations provide an important tool for testing local
helioseismology.

\end{abstract}


\section{Introduction}
Solar 5-min. oscillations are excited by turbulent convection
(downdrafts) in subsurface layers of the Sun. These oscillations
consist of acoustic and surface gravity waves with frequencies in
the range of 2$\div$8 mHz and in a wide range of wave numbers. The
observed oscillations can be used for reconstruction of internal
structure of the Sun by methods of helioseismology. There are
several methods of investigation of the interaction of traveling
acoustic waves with small perturbations to the background state. One
of them is the time-distance approach
\citep{Duvall1993,Kosovichev1996}. The key concept of this method is
measuring and inversion of wave travel times. Propagation of
acoustic waves in this approach is calculated using various
approximations to the wave equations, such as the ray theory or
first Born approximation
\citep{Kosovichev1997,Kosovichev2000,Jensen2001,Couvidat2004,Couvidat2006}.
These approximations have been tested using simple models for point
sources (e.g. \citet{Birch2001,Birch2004}), but not for realistic
solar conditions, e.g. realistic stratification and random
excitation sources. Such tests, which require direct numerical
simulations, are extremely important for validating the inferences
from time-distance helioseismology and other local helioseismology
methods.

There are two main directions in numerical simulations of solar
oscillations and waves. The first one is to use realistic non-linear
simulations of solar convection. In such simulations, the waves are
naturally excited by convective motions. These simulations reproduce
quite well the solar oscillation spectrum \cite{Stein2004}, and have
been used for testing time-distance helioseismology
\cite{Georgobiani2006}. The second approach is based on linearized
Euler equations describing wave propagation for a given background
state. The background state can be taken from non-linear numerical
simulations or by perturbing the standard solar model. In this
paper, we describe a numerical method and initial simulation results
developed for the second approach.

We developed a 3D code which utilizes a realistic physics and
accurately simulates reflection of acoustic waves from the top
boundary. A realistic equation of state which takes into account
partial ionizations and various corrections is used. The equation of
state was calculated by interpolation of the OPAL tables. In Section
2.1 we give a detailed description of the underlying physics. The
main attention is paid to developing a consistent procedure for
obtaining a convectively stable background model and establishing
realistic top boundary conditions based on the Perfectly Match Layer
(PML) method. In Section 2.2 we describe a semi-discrete numerical
scheme of high order which preserves dispersion relations of the
continuous problem. The main attention is paid to developing stable
high-order numerical boundary conditions consistent with the
finite-difference scheme, in which the dispersion relation is
preserved for the inner mesh points of the computational domain. In
Section 3, we compare the numerical and analytical solutions of
various 1D test problems with an isothermal background model to
validate the code, investigate accuracy of the numerical scheme, and
the non-reflecting boundary conditions in a gravitationally
stratified medium. In Section 4, we present results of numerical
three dimensional simulations of acoustic wave field for a standard
solar background model. We used different types of single or
multiple acoustic sources: z-component of force or pressure, point
or distributed with different time dependence. The main goals are to
study properties of solar waves for various excitation sources and
generate artificial wave fields for testing accuracy of the Born and
ray approximations and local helioseismic diagnostics of the solar
interior, currently used for SOHO/MDI and GONG data. The results of
this testing will be presenting in future papers. The numerical
simulations are carried out on parallel supercomputers at NASA Ames
Research center.

\section{Code description}
\subsection{Physical background}
The propagation of adiabatic acoustic waves below the solar
photosphere is described by the following system of linearized Euler
equations:
\begin{equation}\label{LinEuler}
\begin{array}{l}
\displaystyle\ddt{\rho'} + \ddx{}(\rho_0 u') +
\ddy{}(\rho_0 v') + \ddz{}(\rho_0 w')= 0 \vspace{5pt}\\
\displaystyle\ddt{}(\rho_0 u') +\ddx{p'} = 0 \vspace{5pt}\\
\displaystyle\ddt{}(\rho_0 v') +\ddy{p'} = 0 \vspace{5pt}\\
\displaystyle\ddt{}(\rho_0 w') +\ddz{p'} = -g_0\rho',
\end{array}
\end{equation}
where $u',v',w'$ are the perturbations of $x,y,z$ velocity
components, $\rho'$ and  $p'$ are the density and pressure
perturbations correspondingly. Quantities with subscript 0 such as
the pressure $p_0$, the density $\rho_0$, and the gravitational
acceleration $g_0$ correspond to the background reference model and
depend only on radius. Christensen-Dalsgaard's standard solar model
S \citep{Christensen-Dalsgaard1996} with smoothly joined
chromosphere provided by \citet{Vernazza1976} was chosen as such a
model. To close the system we use an adiabatic relation between
Eulerian variations of pressure $p'$ and density $\rho'$
\begin{equation}\label{Adiabat}
p' = a_0^2\rho' - a_0^2\frac{N_0^2}{g_0}(\rho_0\xi_z),
\end{equation}
where $a_0^2=\Gamma_1p_0/\rho_0$ is the square of sound speed,
$\Gamma_1=(\partial\log p/\partial\log\rho)_{ad}$ is the adiabatic
exponent, $N_0$ is the Brunt-V\"{a}is\"{a}l\"{a} frequency, $\xi_z$
is the vertical displacement. So, from the background model we need
only profiles of the sound speed and Brunt-V\"{a}is\"{a}l\"{a}
frequency.

\begin{figure}
\epsscale{1.0}\plotone{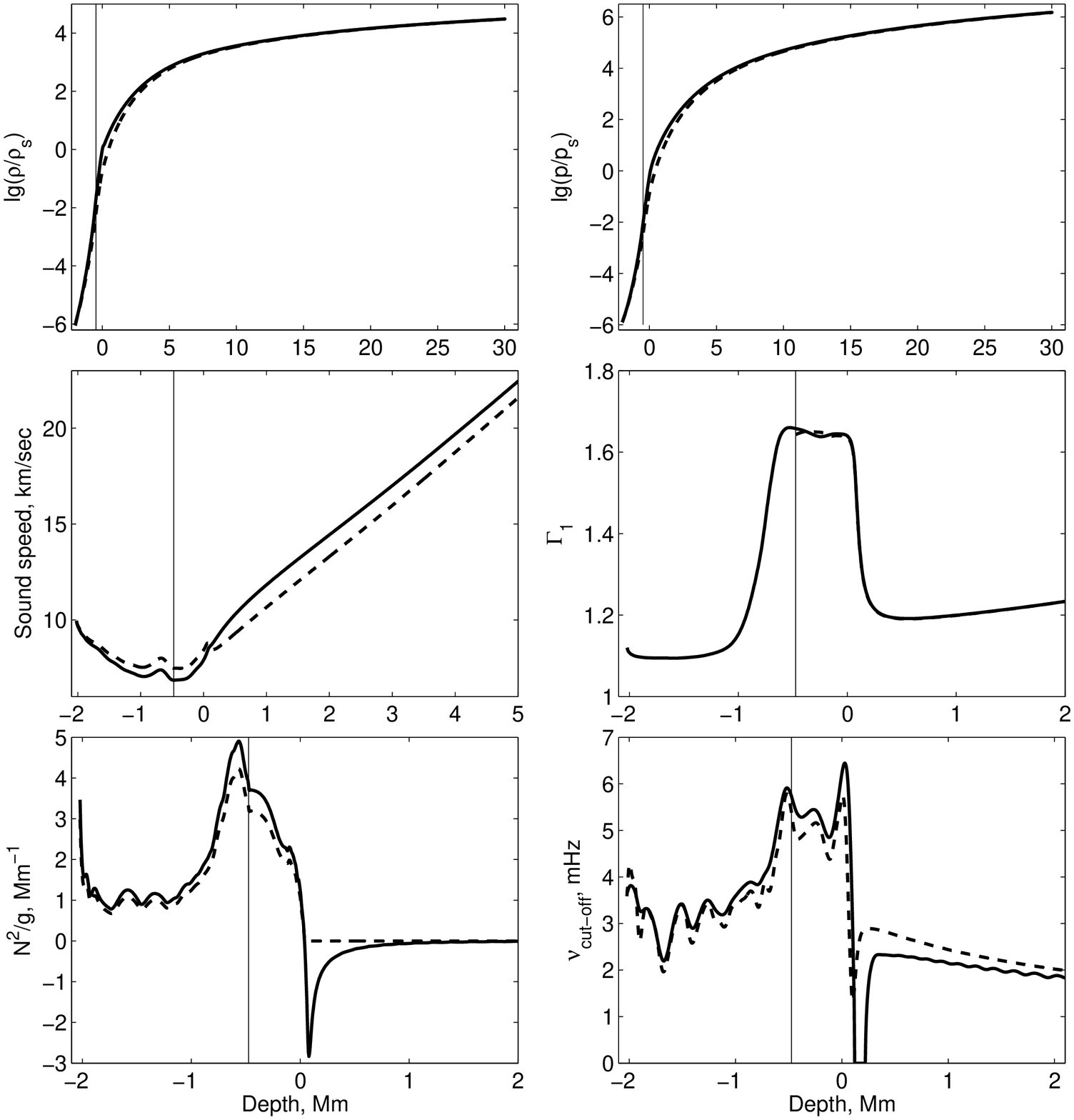} \caption{Vertical profiles of the
density, pressure, sound speed, adiabatic exponent,
Brunt-V\"{a}is\"{a}l\"{a} frequency, and acoustic cut-off frequency.
Solid curves represent profiles for the standard solar model with
smoothly joined chromosphere, dashed ones show the profiles of the
convectively stable modified model. \label{BkgModel_fig}}
\end{figure}

The background state is convectively unstable, especially just below
the photosphere, where the temperature gradient is super-adiabatic
and convective motions are very intense and turbulent. This leads to
the instability of the solution of linear system (\ref{LinEuler}).
Convection instability is developed on a time scale of 30-40 minutes
of solar time. Simulation of the acoustic spectrum of solar
oscillations needs to perform calculations on a time interval of 5-8
hours of solar time, and this instability will badly distort the
result. To make the background model stable against convection we
slightly modified profiles of pressure and density in a thin (500 km
in depth) super-adiabatic layer just below the photosphere. The
condition for stability against convection requires that the square
of Brunt-V\"{a}is\"{a}l\"{a} frequency has to be positive
\begin{equation}\label{BruntVaisala}
N^2(r)=g\left(\frac{1}{\Gamma_1}\frac{\d\log p}{\d r} - \frac{\d\log
\rho}{\d r}\right)>0,
\end{equation}
where $r$ is the distance from the center of the Sun. The profile of
$N^2$ near the solar surface is shown in the bottom left pane of
Figure~\ref{BkgModel_fig} by the solid line. If we replace negative
values by zero (or small positive) value we guarantee a stability of
the modified model against convection. Now we can recalculate the
pressure and density from the modified profile of $N_{mod}^2$.
Combining equation (\ref{BruntVaisala}) with the condition of
hydrostatic equilibrium, we get the following boundary value problem
for $p$ and $\rho$:
\begin{equation}\label{ModifiedModel}
\begin{array}{l}
\displaystyle \frac{1}{\rho}\frac{\d\rho}{\d z} = -\frac{g}{c^2} -
\alpha\frac{N_{mod}^2}{g},\\[9pt]
\displaystyle \frac{\d p}{\d z} =-\rho g,\\[7pt]
0\leq z\leq L,\quad \rho(0)=\rho_0(0),\quad\rho(L)=\rho_0(L),\quad
p(L)=p_0(L),
\end{array}
\end{equation}
where $L$ is the depth of the computational domain, $z$ is the
vertical coordinate which is counted off from the bottom of the
domain. We introduced a free parameter $\alpha$ that has to be
determined, to make the modified model closer to the original one.
We established boundary conditions for the system
(\ref{ModifiedModel}) setting the density at the top and bottom
boundaries equal to solar values. Parameter $\alpha$ does not change
the condition of convective stability if it remains positive.
Introducing of an additional parameter permits us to establish yet
another boundary condition and fix the pressure at the top boundary.
So the recipe to build a convectively stable background model close
to the standard one is the following. We smoothly join the density
profiles of the standard solar model and the chromosphere, obtain
the pressure profile from the condition of hydrostatic equilibrium,
calculate square of the modified Brunt-V\"{a}is\"{a}l\"{a} frequency
profile $N^2_{mod}$, replacing negative values by a zero or small
positive value, and substitute it into the right hand side of
(\ref{ModifiedModel}). Parameter $\alpha$, profiles of density and
pressure of the modified convectively stable model are obtained as a
solution of the eigenvalue problem (\ref{ModifiedModel}). The
profile of $\Gamma_1$, needed for the sound speed profile, is found
from the realistic OPAL equation of state \citep{Rogers1996}.
Vertical profiles of pressure $p_0$, density $\rho_0$, sound speed
$a_0$, square of Brunt-V\"{a}is\"{a}l\"{a} frequency $N^2$,
adiabatic exponent $\Gamma_1$, and acoustic cut-off frequency
\begin{equation}
\omega_c^2 = \frac{a_0^2}{4H^2}\left(1-2\frac{\d H}{\d
r}\right),\quad H^{-1}=-\frac{\d\log\rho_0}{\d r}
\end{equation}
for both models (standard with joined chromosphere and modified
convectively stable) are shown in Figure~\ref{BkgModel_fig}. The
depth of the domain from the photosphere equals 30 Mm, the height of
the chromosphere equals 2 Mm, $\alpha=0.861$. The dashed curves
represent profiles in the convectively stable model, the solid lines
show profiles of the standard solar model. The thin vertical line
marks the position of the fitting point of the chromosphere and the
standard solar model. The background model, obtained with the
procedure described above, is convectively stable and
self-consistent.

\subsection{Numerical algorithm}
The system (\ref{LinEuler}) is written in a conservative form
\begin{equation}\label{DivForm}
\ddt{\mi{q}}+\ddx{\mi{F}(\mi{q})}+
\ddy{\mi{G}(\mi{q})}+\ddz{\mi{H}(\mi{q})}=\mi{S}(\mi{q},t),
\end{equation}
where $\mi{q}=(\rho',\rho_0u',\rho_0v',\rho_0w')^T$ is the vector of
independent variables, $\mi{S}(\mi{q},t)$ is the source term
containing the gravity term and acoustic sources which depend on
time explicitly. The source term does not contain spatial
derivatives. Explicit expressions for components of vectors
$\mi{F}$, $\mi{G}$, and $\mi{H}$ can be easily found from the system
(\ref{LinEuler}). We used a semi-discrete numerical scheme. In
semi-discrete approach the space and time discretization processes
are separated. First the spatial discretization is performed,
leaving the problem continuous in time. The spatial derivatives have
been approximated by the finite difference method reducing the
system of partial differential equations to the system of ordinary
differential equations
\begin{equation}\label{PDE2ODE}
\begin{array}{lcl}
\displaystyle\DDt{\mi{q}_{ikj}}&=&L_{ikj}(\mi{q},t),\\[7pt]
\displaystyle L_{ikj}(\mi{q},t)&=&\displaystyle -\frac{1}{\Delta
x}\sum_{l=-3}^3a_l\mi{F}_{i,k,j+l} -\frac{1}{\Delta
y}\sum_{l=-3}^3a_l\mi{G}_{i,k+l,j}\\[7pt]
&-&\displaystyle\frac{1}{\Delta z}\sum_{l=-3}^3a_l\mi{H}_{i+l,k,j}+
\mi{S}_{ikj}(t)
\end{array}
\end{equation}
which can be solved by any stable time advancing method. We used
four stage, 3rd order strong-stability-preserving Runge-Kutta method
\citep{Shu2002} with Courant number $c=2$:
\begin{equation}\label{SSP_RK34}
\begin{array}{l}
\displaystyle\mi{q}^{(1)}=\mi{q}^{(n)}+\frac{1}{2}\Delta
t\, L(\mi{q}^{(n)},t^n),\\[7pt]
\displaystyle\mi{q}^{(2)}=\mi{q}^{(1)}+\frac{1}{2}\Delta
t\, L(\mi{q}^{(1)},t^n+\frac{1}{2}\Delta t),\\[7pt]
\displaystyle\mi{q}^{(3)}=\frac{2}{3}\mi{q}^{(n)}
+\frac{1}{3}\mi{q}^{(2)}+\frac{1}{6}\Delta t\,
L(\mi{q}^{(2)},t^n+\Delta t),\\[7pt]
\displaystyle\mi{q}^{(n+1)}=\mi{q}^{(3)}+\frac{1}{2}\Delta t\,
L(\mi{q}^{(3)},t^n+\frac{1}{2}\Delta t).
\end{array}
\end{equation}
The source term $\mi{S}(\mi{q},t)$ which depends on time explicitly
does not require a special treatment.

\begin{figure}
\epsscale{1.0} \plotone{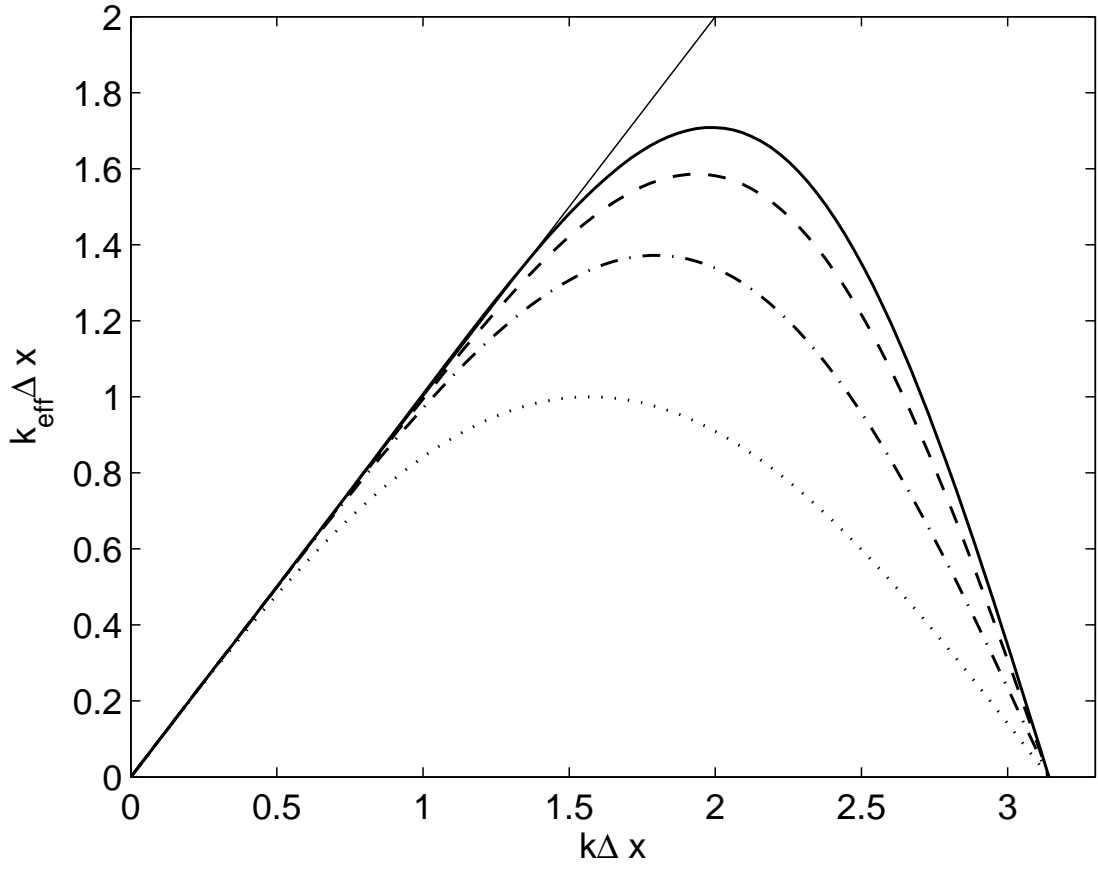} \caption{Effective wave number
$k_{eff}\Delta x$ versus $k\Delta x$ for different numerical
schemes. Dotted, dash-dotted, dashed, and solid curves represent
classic 2nd, 4th, 6th, and DRP 4th order schemes correspondingly.
\label{Wavenum_fig}}
\end{figure}

High-order dispersion-relation-preserving (DRP) scheme developed by
\citet{Tam1993} was used for spatial discretization. Coefficients
$a_l$ of the finite difference scheme
\begin{equation}\label{ddx_approx}
\left.\ddx{f}\right|_j\simeq\frac{1}{\Delta
x}\sum_{l=-3}^3a_lf_{j+l}=\frac{1}{\Delta x}\sum_{l=-3}^3a_l
f(x_j+l\Delta x)
\end{equation}
are chosen from the requirement that the difference between Fourier
transform of the numerical scheme and Fourier transform of the
spatial derivative has to be minimal. Taking the Fourier transform
of both sides of (\ref{ddx_approx}) one can get the effective wave
number $k_{eff}$ of the Fourier transform of the numerical scheme
(\ref{ddx_approx})
\begin{equation}\label{EffWaveNumber}
k_{eff}=-\frac{i}{\Delta x}\sum_{l=-3}^3a_le^{ilk\Delta x}.
\end{equation}
An assumption that the integral error $E$ of Fourier transform of
the finite difference scheme (\ref{ddx_approx}) is minimal for waves
with wavelength $\lambda\geq4\Delta x$ leads to the following
equation
\begin{equation}\label{FourierError}
\dpp{E}{a_j}=0,\quad j=-1,1,\quad E=\int_{-\pi/2}^{\pi/2}|k\Delta x
- k_{eff}\Delta x|^2 \,\d(k\Delta x),
\end{equation}
or in the explicit form
\begin{equation}
\begin{array}{l}
\displaystyle -2-\frac{2a_{-2}}{3}+2a_0+\pi a_1+2a_2 = 0,\vspace{5pt}\\
\displaystyle 2+2a_{-2}+\pi a_{-1}+2a_0-\frac{2a_2}{3} = 0.
\end{array}
\end{equation}
The rest five equations
\begin{equation}
\sum_{j=-3}^3 a_j = 0, \sum_{j=-3}^3 ja_j = 1, \sum_{j=-3}^3 j^2a_j
= 0, \sum_{j=-3}^3 j^3a_j = 0, \sum_{j=-3}^3 j^4a_j = 0
\end{equation}
are obtained from a requirement that the numerical scheme
(\ref{ddx_approx}) approximates a spatial derivative with the 4th
order. Now calculation of the coefficients of 7-dot symmetrical
stencil is straightforward. The explicit expressions for the
coefficients are the following:
\begin{equation}\label{SpatialCofs}
\left\{
\begin{array}{l}
\displaystyle a_0 = 0 \vspace{5pt}\\
\displaystyle a_{\pm 1} = \pm\frac{12}{15\pi-32} \vspace{5pt}\\
\displaystyle a_{\pm 2} = \mp\frac{96-27\pi}{60\pi-128} \vspace{5pt}\\
\displaystyle a_{\pm 3} = \pm\frac{20-6\pi}{45\pi-96}.
\end{array}
\right.
\end{equation}
The plots of numerical wave number $k_{eff}\Delta x$ versus $k\Delta
x$ for different numerical schemes are shown in
Figure~\ref{Wavenum_fig}. Dotted, dash-dotted, dashed, and solid
curves represent classic 2nd, 4th, 6th, and DRP 4th order schemes
correspondingly. One can see that the 4th-order DRP scheme describes
short waves more accurately than the classic 6th-order scheme.

\begin{figure}
\epsscale{1.0} \plotone{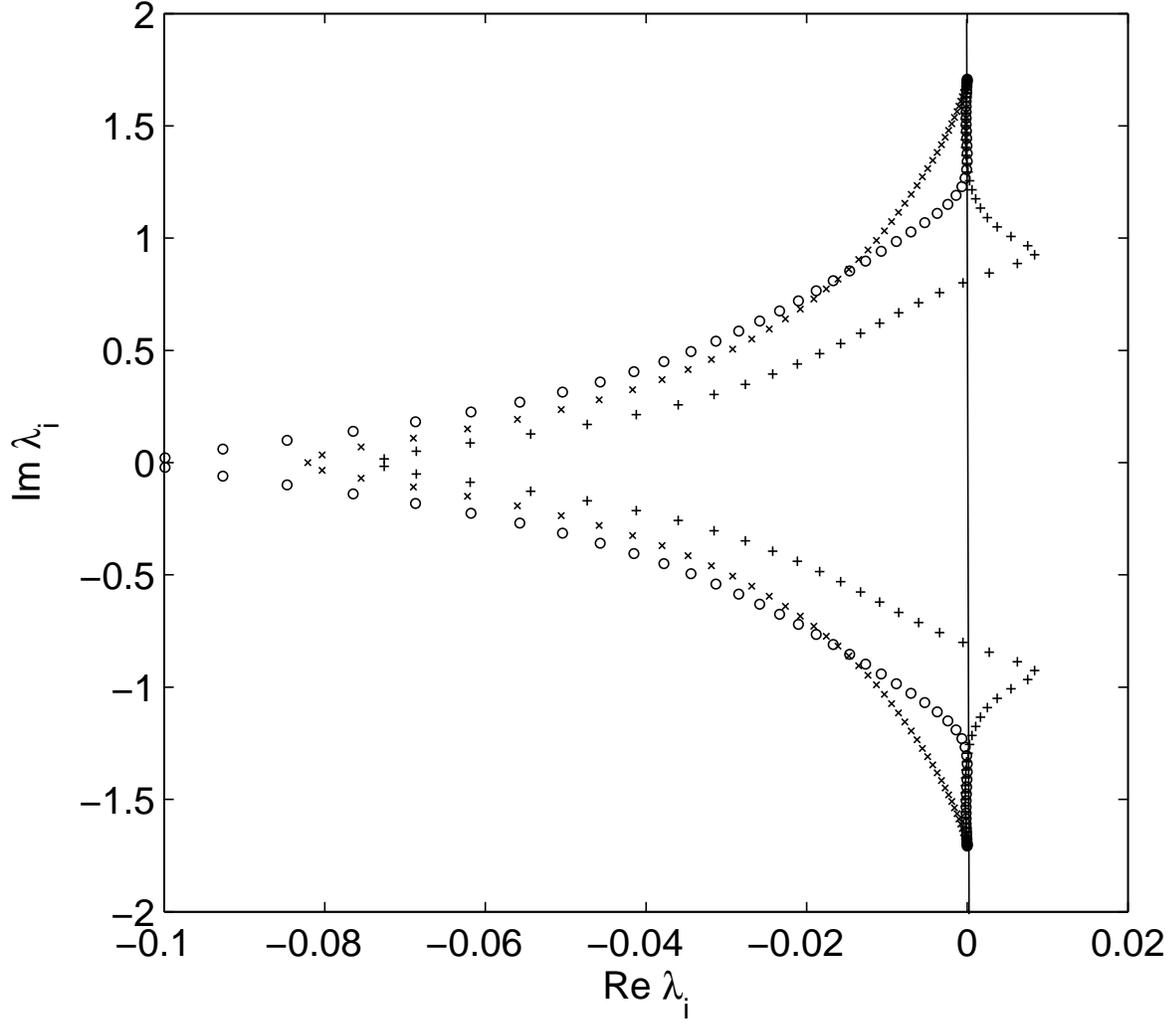} \caption{Eigenvalues of the DRP
spatial discretization operator for scalar advection equation on
complex plane for different choices of the coefficient $p_{23}$.
Symbols plus correspond to the scheme $p_{23}=1/30,\:p_{33}=31/32$
which does not exhibit an asymptotic stability. Circles and crosses
represent choices of $p_{23}=1/80,\,-1/10$ and $p_{33}=125/128,\:
65/64$ correspondingly. Both schemes are asymptotically stable.
\label{Eigs_fig}}
\end{figure}

Waves with wavelength less than $4\Delta x$ are not resolved by the
numerical scheme (\ref{ddx_approx}). They lead to point-to-point
oscillations of the solution that can cause a numerical instability.
Such waves have to be filtered out. We used the following digital
filter of the 6th-order to eliminate unresolved short wave component
from the solution:
\begin{equation}\label{filter}
f_{sm}(x) = f(x)-\sigma_f D(x) = f(x) - \sigma_f\sum_{m=-3}^3
d_mf(x+m\Delta x),
\end{equation}
where $f$ is the original grid function, $f_{sm}$ is the filtered
grid function, $D(x)$ is the damping function, $\sigma_f$ is the
constant between 0 and 1, determining the filter strength. The
frequency response function $G(k)$ of the filter relates the Fourier
images of the original $\tilde{f}$ and filtered $\tilde{f}_{sm}$
grid functions as follows $\tilde{f}_{sm}(k)=G(k)\tilde{f}(k)$. In
this paper, the coefficients $d_m$ of the digital filter have been
chosen in such a way that
\begin{equation}
G(k\Delta x) = 1 - \sum_{m=-3}^3d_me^{imk\Delta x}=
1-\sin^{6}\left(\frac{k\Delta x}{2}\right).
\end{equation}
Coefficients $d_m$ of the digital filter are symmetric:
\begin{equation}
\begin{array}{l}
\displaystyle d_0=\frac{5}{16},\quad d_1=d_{-1}=-\frac{15}{64},\quad
d_2=d_{-2}=\frac{3}{32},\quad d_3=d_{-3}=-\frac{1}{64}.
\end{array}
\end{equation}
Using the technique proposed by \citet{Carpenter1993} we have found
a stable 3rd-order boundary closure of the explicit
dispersion-relation-preserving inner scheme (\ref{ddx_approx}) with
coefficients given by (\ref{SpatialCofs}). Obtained boundary closure
has summation-by-parts properties. This approach is based on the
implicit Pad\'{e} approximation of spatial derivatives near the
boundaries
\begin{equation}\label{PadeDeriv}
\mi{P} \dpp{\mi{q}}{x} = \mi{Qq},
\end{equation}
where matrices $\mi{P}$ and $\mi{Q}$ satisfy the following
conditions:
\begin{enumerate}
\item $\mi{P}$ is symmetric non-singular matrix ($\mi{P} = \mi{P}^T$),
\item $\mi{P}$ is positive-definite matrix
($\mi{U}^T\mi{PU}>0$ for $\forall\;\mi{U}$),
\item $\mi{Q}$ is almost skew-symmetric matrix, except corner top left and
bottom right elements
($\mi{Q}+\mi{Q}^T=|q_{0,0}|\mbox{diag}(-1,0,\ldots,1)$)
\item $q_{N,N}>0,\quad q_{0,0}=-q_{N,N}$.
\end{enumerate}
Taking into account these properties, one can write explicitly the
top left corners of matrices $\mi{P}$ and $\mi{Q}$:
\begin{equation}\label{PQ_43}
\begin{array}{l}
\mi{Q}=\left(
\begin{array}{ccccccccc}
q_{00}  & q_{01}  & q_{02} & q_{03} & 0 & 0 & 0 & 0 &\\
-q_{01} & 0       & q_{12} & q_{13} & a_3  & 0 & 0 & 0 &\\
-q_{02} & -q_{12} & 0      & q_{23} & a_2 & a_3 & 0 & 0 &\cdots\\
-q_{03} & -q_{13} & -q_{23} & 0 & a_1 & a_2 & a_3 & 0 &\\
0 & -a_3 & -a_2 & -a_1 & 0 & a_1
&a_2 & a_3 &\\
& & & \vdots & & & & & \ddots\\
\end{array} \right),\\
\mi{P}=\left(
\begin{array}{cccccc}
p_{00} & p_{01} & p_{02} & p_{03} & 0 &\\
p_{01} & p_{11} & p_{12} & p_{13} & 0 &\\
p_{02} & p_{12} & p_{22} & p_{23} & 0 & \cdots\\
p_{03} & p_{13} & p_{23} & p_{33} & 0 &\\
0      & 0      & 0      & 0      & 1 &\\
& & \vdots & & & \ddots
\end{array} \right),
\end{array}
\end{equation}
where coefficients $a_i$ of the inner scheme are defined in
(\ref{SpatialCofs}). Expanding the left and right-hand sides of
(\ref{PadeDeriv}) in Taylor series at the top boundary and equating
terms of the same order of $\Delta x$, one can obtain a system of
linear equations for coefficients $p_{ij}$ and $q_{ij}$. Not all of
these equations are independent, so the solution depends on two free
parameters $p_{33}$ and $p_{23}$:
\begin{equation}
\begin{array}{l}
\displaystyle p_{00}=-\frac{83}{108}+p_{33},\quad
p_{11}=-8p_{23}-15p_{33}-\frac{34016-14973\pi}{54(15\pi-32)},\vspace{5pt}\\
\displaystyle p_{22}=\frac{1727}{108}-8p_{23}-15p_{33},\quad
p_{01}=p_{23}-\frac{8(6\pi-29)}{27(15\pi-32)},\vspace{5pt}\\
\displaystyle p_{02}=3p_{33}-\frac{10929\pi-24352}{216(15\pi-32)},
\quad p_{03}=-p_{23}-4p_{33}-\frac{58528-26949\pi}{432(15\pi-32)},\vspace{5pt}\\
\displaystyle
p_{12}=7p_{23}+12p_{33}-\frac{84813\pi-188192}{432(15\pi-32)},
\quad p_{13}=3p_{33}-\frac{8691\pi-17504}{216(15\pi-32)},\vspace{5pt}\\
\displaystyle q_{00}=-\frac{1}{2}, \quad
q_{01}=-2p_{23}-6p_{33}-\frac{212960-95451\pi}{864(15\pi-32)},\vspace{5pt}\\
\displaystyle
q_{02}=4p_{23}+12p_{33}-\frac{42537\pi-93856}{216(15\pi-32)}, \quad
q_{03}=-2p_{23}-6p_{33}-\frac{176288-81177\pi}{864(15\pi-32)},\vspace{5pt}\\
\displaystyle
q_{12}=-6p_{23}-18p_{33}-\frac{180896-83661\pi}{288(15\pi-32)},
\quad
q_{13}=4p_{23}+12p_{33}-\frac{38451\pi-80992}{216(15\pi-32)},\vspace{5pt}\\
\displaystyle
q_{23}=-2p_{23}-6p_{33}-\frac{152288-76731\pi}{864(15\pi-32)}.
\end{array}
\end{equation}
To satisfy a condition of positive definiteness it is sufficient to
choose matrix elements $p_{33}$ and $p_{23}$ in such a way, that the
signs of coefficients of a characteristic polynomial alternate.
However, this property does not guarantee a boundedness of the
solution for all times, which is called asymptotic stability. To
make a solution be bounded for all times, all eigenvalues of the
spatial discretization operator $L_{ikj}$ from Eq.(\ref{PDE2ODE}),
incorporated with the boundary conditions, must have non-positive
real parts. Details of this procedure can be found in
\citep{Carpenter1993}. Due to complexity of the original 3D problem,
we have tested a stability of the scheme on 1D advection problem.
Distribution of eigenvalues of the DRP spatial discretization
operator in the complex plane for different choices of the pairs of
coefficients ($p_{23}$, $p_{33}$) is shown in Fig.~\ref{Eigs_fig}.
Symbols plus correspond to the scheme $p_{23}=1/30,\:p_{33}=31/32$,
which does not exhibit asymptotic stability. Circles and crosses
represent choices (1/80, 125/128) and (-1/10, 65/64) of coefficients
$(p_{23},p_{33})$ correspondingly. Both schemes are asymptotically
stable.

Acoustic sources have been added to the right-hand side of equations
(\ref{LinEuler}). We used sources of two types. If we add a scalar
function $\Phi(x,y,z,t)$ to the right-hand side of z-component of
momentum equation, this term can be combined with the gravity term
and interpreted as a source of z-component of force. If we add a
gradient of a scalar function $\nabla\Phi(x,y,z,t)$ to the
right-hand side of all momentum equations, these terms can be
combined with components of the pressure gradient and interpreted as
a pressure source. Acoustic sources are spatially localized and have
finite lifetime. Spatial dependence is given by Gaussian spherically
symmetric function with semi-width of 2-3 grid nodes. We
experimented with two different time dependencies of acoustic
sources: one period of sin function $\sin[\omega(t-t_0)],\; t_0\leq
t\leq t_0+2\pi/\omega$ and Ricker's wavelet $(1-2x^2)e^{-x^2},\; x=
[\omega (t-t_0)/2-\pi],\; t_0\leq~t\leq~t_0+4\pi/\omega$. Such time
dependencies were chosen because such sources are not monochromatic
and have spectral power localized around central frequency
$\omega/2\pi$, but spectral power is not too spread out. Single or
multiple acoustic sources can be added to the right hand side of
equations. Multiple sources are randomly distributed at some depth
(in our case it was 350 km) and initiated independently at arbitrary
moments of time. Amplitudes and frequencies are randomly distributed
on intervals [0, 1] and [2 mHz, 8 mHz] correspondingly. Thickness of
both (top and bottom) PMLs equals 5 grid nodes.

Calculation of the acoustic spectrum requires long term simulations,
so besides the asymptotic stability we have to prevent a spurious
reflection of acoustic waves from the boundaries back to the
computational domain. In this paper we follow \citet{Hu1996}, who
proposed a procedure to construct the PML for the Euler equations.
It can be proofed that for a homogeneous medium and uniform mean
flow without gravity PML absorbs waves without reflection for any
angle of incidence and frequency. We set non-reflecting boundary
conditions based on the PML at the top and bottom boundaries of the
domain. The lateral boundary conditions are periodic. Inside the PML
independent variables $\mi{q}$ are split into sub-components
$\mi{q}_1, \mi{q}_2, \mi{q}_3$ such that $\mi{q}=\mi{q}_1 + \mi{q}_2
+ \mi{q}_3$. Thus, in the PML 3D system (\ref{LinEuler}) is split
into 1D+1D+1D system of coupled locally one dimensional equations
\begin{equation}
\begin{array}{l}
\displaystyle \ddt{\mi{q}_1} +
\ddx{\mi{F}(\mi{q})} = 0,\\[7pt]
\displaystyle \dpp{\mi{q}_2}{t} +
\ddy{\mi{G}(\mi{q})} = 0,\\[7pt]
\displaystyle \dpp{\mi{q}_3}{t} + \ddz{\mi{H}(\mi{q})} =
\mi{S}(\mi{q},t) - \sigma_z\mi{q}_3,
\end{array}
\end{equation}
where $\Delta t\,\sigma_z=0.05+\sigma_{max}(Z/D)^2$ is the damping
factor, $Z$ is the vertical coordinate inside the PML counted off
from the interface of the PML with the inner region, D is the depth
of the PML. Values of $\sigma_{max}$ at the top and bottom
boundaries are 0.3 and 1.0 correspondingly. In the paper of F.Q. Hu
a quadratic dependence of $\sigma$ on the coordinate $Z$ is used. We
were forced to add a small constant term to stabilize the PML in the
presence of gravity. It is important to note, that vectors $\mi{F}$,
$\mi{G}$, and $\mi{H}$ depend only on unsplit variable $\mi{q}$.
Although $\mi{q}_1$, $\mi{q}_2$, and $\mi{q}_3$ are not defined
outside the PML, the variable $\mi{q}$, which is used for
calculation of the spatial derivatives, is defined everywhere in the
computational domain. Hence, inside the PML near the interface with
the inner region we can use the same centered stencil as for the
inner points. Near the top and bottom boundaries the implicit Pade
approximation (\ref{PadeDeriv}) is used which guarantees numerical
stability of the scheme. We smoothly joined the chromospheric model
provided by \citet{Vernazza1976} with the top of the standard solar
model by Christensen Dalsgaard and established the top
non-reflecting boundary condition based on the PML in the
chromosphere above the temperature minimum. This simulates a
realistic situation when not all waves are reflected by the
photosphere. Waves with frequencies higher than the acoustic cut-off
frequency pass through the photosphere and will be absorbed by the
PML layer.

\section{Numerical examples}
\begin{figure}
\epsscale{1.0} \plotone{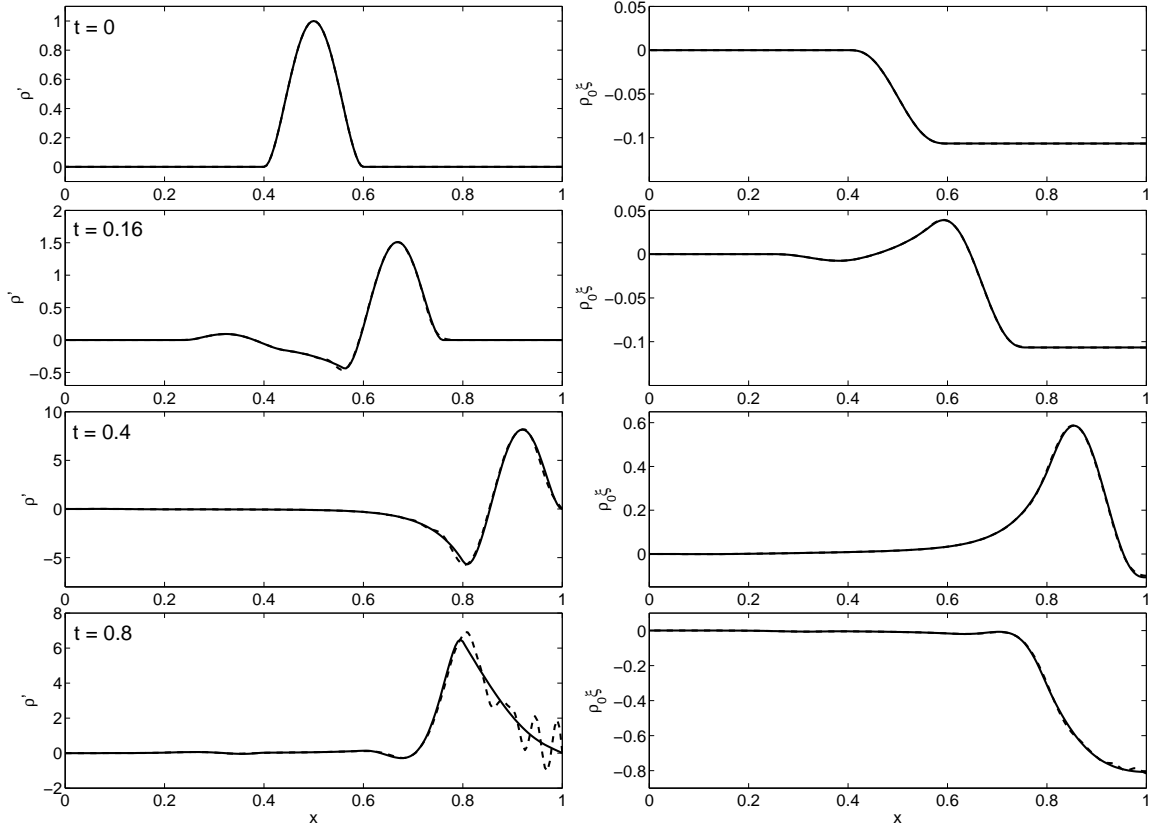} \caption{Solution of the IBVP
(\ref{IBVP_1D}) for the isothermal hydrostatic background model.
Density and displacement are shown on the left and right panes
correspondingly. Solid curve represents the exact solution, dashed
one shows the classic 2/1 numerical solution. The high order DRP 4/3
numerical solution is indistinguishable from the exact one. After
hitting the boundary classic 2/1 solution changes the global order
of accuracy to 1, because boundary conditions are realized only with
the 1st order of accuracy. \label{IBVP_fig}}
\end{figure}

For validation of the code we chose 1D initial boundary value
problem (IBVP) for linearized Euler equations with constant gravity
$g_0=const$:
\begin{equation}\label{IBVP_1D}
\begin{array}{l}
\displaystyle \ddt{\rho'}+\ddx{}(\rho_0u')=0,\\[9pt]
\displaystyle \ddt{}(\rho_0u')+\ddx{p'}=g_0\rho',\\[9pt]
\displaystyle \ddt{}(\rho_0\xi)=\rho_0u',\\[9pt]
\displaystyle p'=a_0^2\rho'+(\gamma-1)g_0(\rho_0\xi),
\end{array}
\begin{array}{l}
0\leq x\leq 1,\quad t\geq 0,\\[9pt]
\rho'(0,t)=\rho'(1,t)=0,\\[9pt]
\rho'(x,0)=f(x),\quad \rho_0(x)u'(x,0)=0,\\[9pt]
\displaystyle \rho_0(x)\xi(x,0)=-\int\limits_0^xf(\eta)\;\d\eta,
\end{array}
\end{equation}
where $\xi$ is the displacement. We need to know $\xi$ to calculate
the Eulerian perturbation of pressure $p'$. Actually, the
combination $\rho_0\xi$ is used as a variable, so we group them
together in equations. Initial conditions for $\xi$ must be
consistent with the initial conditions for $\rho'$. They are related
by the continuity equation. Waves are adiabatic, the background
model $p_0, \rho_0$ is hydrostatic and isothermal
$p_0/\rho_0=const$. The last equation in the system (\ref{IBVP_1D})
is the adiabatic relation (\ref{Adiabat}) written for the isothermal
background model. Variable $x$ represents here the depth from the
surface. The system (\ref{IBVP_1D}) is written in the same
conservative form as the original system (\ref{LinEuler}). This
problem was chosen for testing the code because it shows all
characteristic behavior of the realistic solution and yet not too
complicated and can be solved analytically. Formally, the variable
$\rho_0\xi$ can be eliminated using the continuity equation in 1D,
and system (\ref{IBVP_1D}) can be reduced to the system of two
equations. However, in 3D $\rho_0\xi$ cannot be eliminated. To have
the test example as close to the real case as possible, we left this
variable and solved numerically the full system (\ref{IBVP_1D}).
Analytical solution of these equations can be obtained by the method
of separation of variables
\begin{equation}\label{IBVP_rhoxi}
\begin{array}{l}
\displaystyle \rho'(x,t)=e^{x/2H}\sum_{n=1}^\infty A_n
\sin\pi n x\cos \lambda_na_0 t,\\[9pt]
\displaystyle \xi(x,t)=e^{-x/2H}\sum_{n=1}^\infty
B_n(\sin\pi nx-2\pi nH\cos\pi nx) \cos \lambda_na_0t,\\[9pt]
\displaystyle A_n=2\int_0^1 f(\eta)e^{-\eta/2H}\sin\pi n\eta
\d\eta,\quad B_n = -\frac{2HA_n}{1+4\pi^2n^2H^2},\\[9pt]
\displaystyle \lambda_n=\sqrt{\frac{1}{4H^2}+\pi^2n^2},\quad
H^{-1}=\frac{\gamma g_0}{a_0^2}.
\end{array}
\end{equation}
The following distribution of density perturbation $f(x)$ was chosen
as the initial condition for $\rho'$:
\begin{equation}\label{IBVP_rhoini}
f(x)=\left\{\begin{array}{l}
10^4[(x-0.5)^2-0.001]^2\quad \mbox{if}\;\; 0.4\leq x\leq 0.6 \\
0
\end{array}
\right.
\end{equation}
Solution of the initial boundary value problem (\ref{IBVP_1D}) for
different moments of time with the initial density distribution
given by (\ref{IBVP_rhoini}) and parameters $a_0=1$, $\gamma=5/3$,
$g_0=10$, $\Delta t=2\cdot10^{-3}$, $N=200$ (number of grid nodes)
is shown in Figure \ref{IBVP_fig}. The left column represents the
density perturbation, the right one shows the vertical displacement.
The solid curve is the exact solution (\ref{IBVP_rhoxi}). The dashed
curve represents the low-order (classic 2/1) numerical solution
which uses the 2nd-order classic central difference approximation of
spatial derivatives for inner points with the one sided 1st-order
scheme at the boundaries. The high-order numerical solution is
indistinguishable from the exact one. It uses the DRP 4/3 scheme
(dispersion-relation-preserving spatial discretization of the 4th
order for inner points with the stable boundary closure of the 3rd
order consistent with the inner scheme). The bottom panels give the
profiles of density and displacement after reflection from the
bottom boundary. One can see, that the second order solution
approximates the exact one well enough before the wave hits the
boundary. After this the accuracy of the solution switches from the
second to the first order. Solution becomes too dispersive which
causes nonphysical oscillations. The high-order solution based on
DRP 4/3 scheme reproduces the exact solution well even after
$30000\div40000$ iterations and 20$\div$30 reflections from
boundaries. This test shows that the high-order DRP numerical scheme
does not introduce a noticeable damping or dispersion even on big
intervals of integration. These simulations also test an accuracy
and stability of the numerical boundary conditions.

\begin{figure}
\epsscale{1.0} \plotone{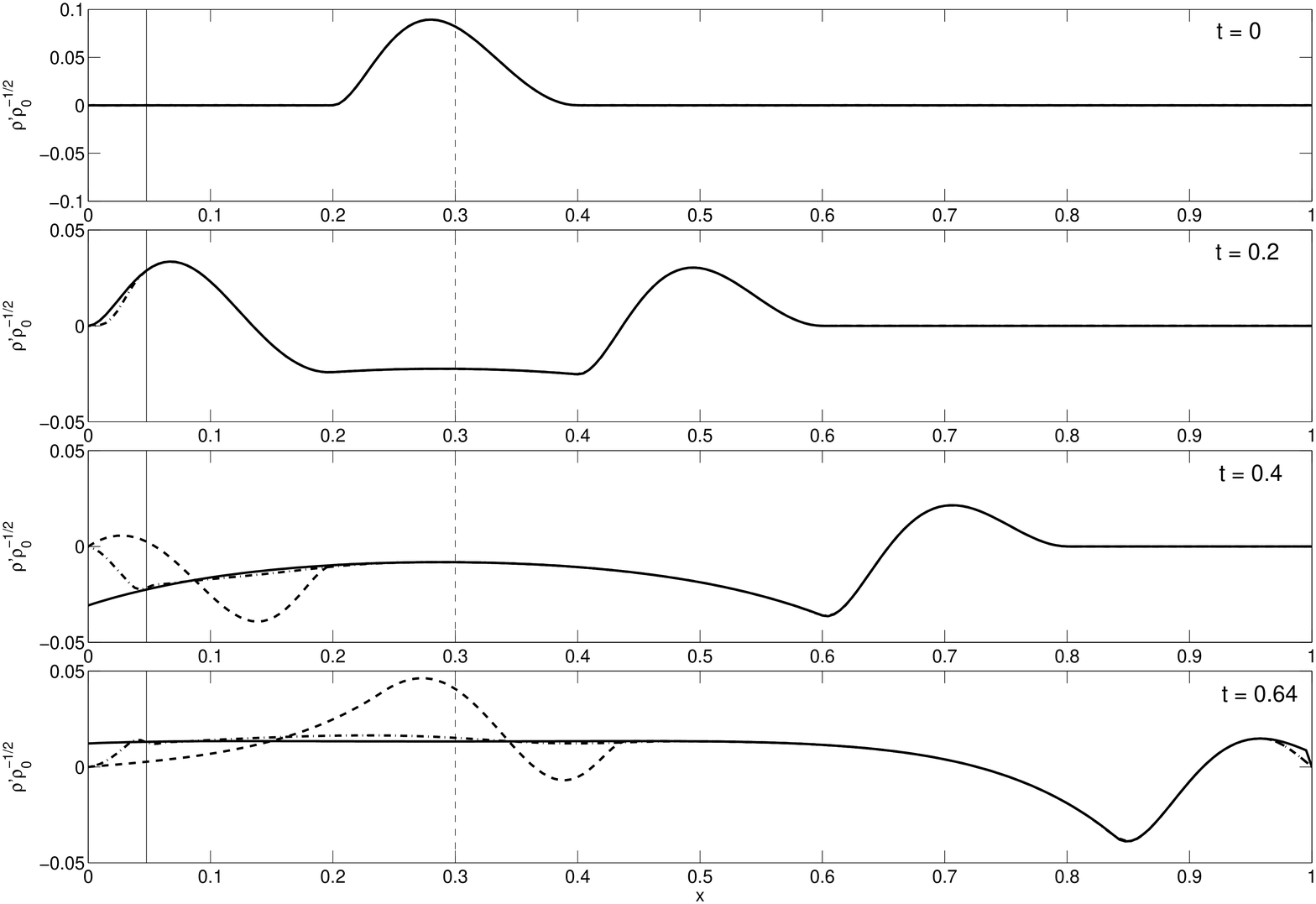} \caption{Density perturbation with
removed exponential factor $\rho'/\sqrt{\rho_0}$ is plotted. The
solid line represents the exact solution for infinite interval
(\ref{Inf_1D}), the dash-dotted line represents the numerical
solution with the PML at the top boundary, and the the dashed line
represents the exact solution (\ref{IBVP_rhoxi}) for reflecting left
boundary. The vertical solid and dashed lines mark positions of the
PML interface and initial perturbation correspondingly.
\label{PML_fig}}
\end{figure}

To test the efficiency of the PML for non-uniform isothermal
background model we compared the numerical solution of
(\ref{IBVP_1D}) with the PML established at the top boundary with
the exact solution of the same problem for infinite interval
$-\infty \leq x\leq\infty$:
\begin{eqnarray}\label{Inf_1D}
\rho'(x,t) &=& \frac{1}{2}f(x+a_0t)e^{-a_0t/2H} +
\frac{1}{2}f(x-a_0t)e^{a_0t/2H} - \nonumber\\
&
&\frac{a_0t}{4H}\;e^{x/2H}\int\limits_{x-a_0t}^{x+a_0t}e^{-\eta/2H}
\frac{J_1(\sqrt{a_0^2t^2-(x-\eta)^2}/2H)}{\sqrt{a_0^2t^2-(x-\eta)^2}}
f(\eta)\;\d\eta.
\end{eqnarray}
Bottom boundary condition for the numerical solution remains
reflecting, because the bottom PML is inconsistent with the initial
conditions for $\xi(x,t)$. At the bottom boundary in the initial
moment of time $\xi(1,0)<0$ (see the top right panel of
Figure~\ref{IBVP_fig}). If we established PML at the bottom, it
would damp $\xi$ to zero value, generating non-physical
perturbations near the bottom boundary which corrupt the solution.
The analytical solution (\ref{Inf_1D}) does not contain reflected
waves, because all initial perturbations propagate to infinity. This
solution can be used as a reference solution for determining the
damping properties of the top PML. Results are shown in
Figure~\ref{PML_fig} at the moments of time $t=$ 0, 0.2, 0.4, and
0.64. The value $\rho'/\sqrt{\rho_0}$ (density perturbation with
removed exponential factor) is plotted. The solid line represents
the exact solution (\ref{Inf_1D}), the dash-dotted line represents
the numerical solution with PML at the top boundary, and the the
dashed line represents the exact solution (\ref{IBVP_rhoxi}) for the
reflecting top boundary. The solid vertical line marks position of
the interface between the top PML and inner region. The dashed
vertical line shows position of the initial perturbation. The top
PML reduces the amplitude of reflected wave by factor 20$\div$40.

\begin{figure}
\epsscale{1.0} \plotone{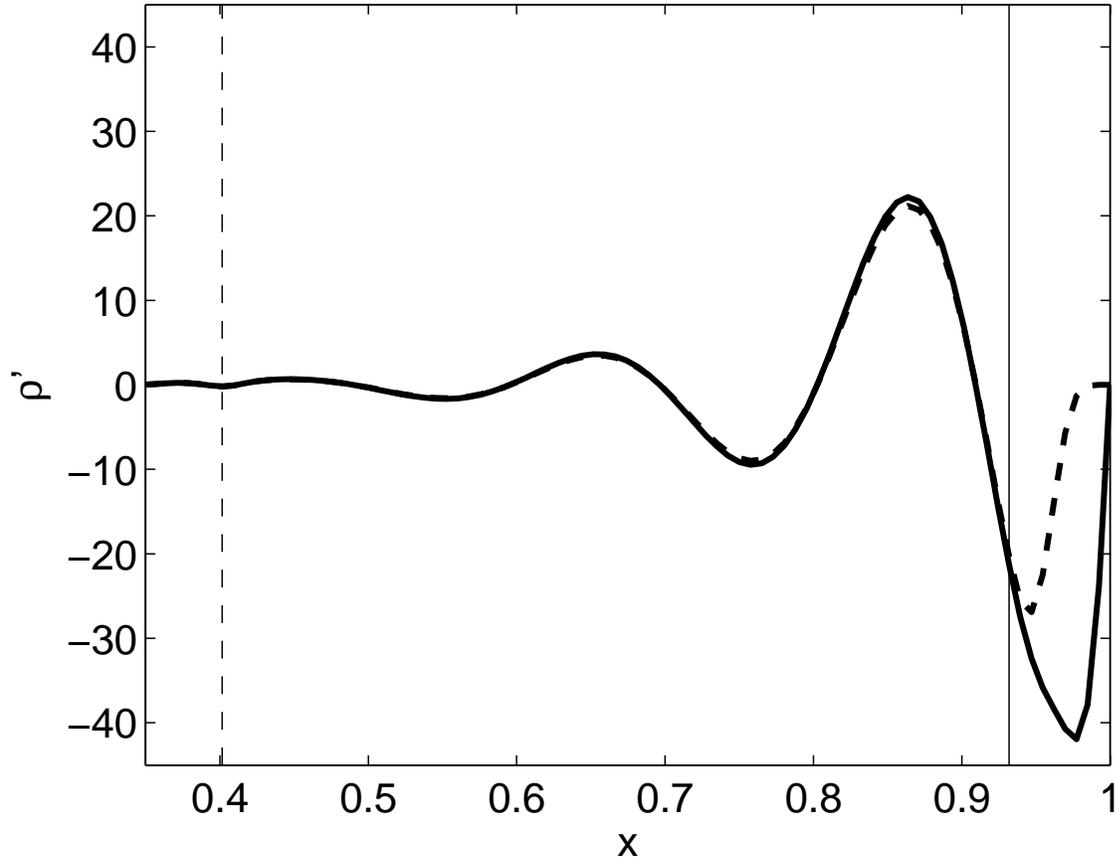} \caption{Solution of the IBVP
(\ref{SRC_1D}) for the isothermal hydrostatic background model in
presence of the source. The solid curve represents the exact
solution with zero boundary conditions for $\rho'$. The dashed line
represents DRP 4/3 numerical solution of (\ref{SRC_1D}) with
non-reflecting top and bottom boundaries. Numerical solution is
damped effectively by the absorbing layer. \label{SRC_1D_fig}}
\end{figure}

Our original 3D system contains acoustic sources explicitly
depending on time in the right hand side of the momentum equations.
We have tested the code in presence of sources on the same problem
(\ref{IBVP_1D}) with the pressure source term
\begin{equation}\label{SRC_1D}
\begin{array}{l}
\displaystyle \ddt{\rho'}+\ddx{}(\rho_0u')=0,\\[9pt]
\displaystyle \ddt{}(\rho_0u')+\ddx{p'}=g_0\rho'-\ddx{\Phi(x,t)},\\[9pt]
\displaystyle \ddt{}(\rho_0\xi)=\rho_0u',\\[9pt]
\displaystyle p'=a_0^2\rho'+(\gamma-1)g_0(\rho_0\xi),
\end{array}
\begin{array}{l}
0\leq x\leq 1,\quad t\geq 0,\\[9pt]
\rho'(0,t)=\rho'(1,t)=0,\\[9pt]
\rho'(x,0)=0,\quad \rho_0(x)u'(x,0)=0,\\[9pt]
\displaystyle \rho_0(x)\xi(x,0)=0.
\end{array}
\end{equation}
For test purposes we chose gaussian shaped harmonic source function
as follows
\begin{equation}\label{SRC_func}
\Phi(x,t)=e^{-\textstyle
\left(\frac{x-h_{src}}{\sigma}\right)^2}\sin(\omega_0 t),
\end{equation}
The system (\ref{SRC_1D}) can be solved analytically by quadratures
\begin{equation}
\begin{array}{l}
\displaystyle\rho'(x,t) =
\int\limits_0^t\!\!\int\limits_0^1\frac{\partial^2\Phi(\eta,\tau)}{\partial\eta^2}
G(x,\eta,t-\tau)\; \d\eta\d\tau,\\[15pt]
\displaystyle G(x,\eta,\tau) = 2\;e^{(x-\eta)/2H} \sum_{n=1}^\infty
\frac{\sin\lambda_na_0\tau}{\lambda_na_0} \sin\pi nx \sin\pi
n\eta,\\[15pt]
\displaystyle \lambda_n=\sqrt{\frac{1}{4H^2}+\pi^2 n^2}.
\end{array}
\end{equation}
The results of numerical simulations with the source function given
by (\ref{SRC_func}) and parameters $N=120$, $\Delta
t=2\cdot10^{-3}$, $a_0=1$, $\gamma=5/3$, $g_0=10$, $h_{src}=0.4$,
$\omega_0=10\pi$, $\sigma=0.0178$ are shown in
Figure~\ref{SRC_1D_fig}. The non-reflecting boundary conditions are
established on the top and bottom boundaries for numerical solution.
The solid curve represents the exact solution of (\ref{SRC_1D_fig})
with zero boundary conditions for $\rho'$ established at $x=0$ and
$x=1$. The dashed line represents DRP 4/3 numerical solution of
(\ref{SRC_1D_fig}) with non-reflecting top and bottom boundaries.
The vertical dashed line marks the position of the source. The
vertical solid line shows the position of the interface between the
inner region of the computational domain and the non-reflecting PML.
The numerical solution reproduces the exact one well in the inner
region, and is effectively damped by the absorbing layer, preventing
unwanted reflection from the bottom boundary.

\begin{figure}
\epsscale{1.0} \plotone{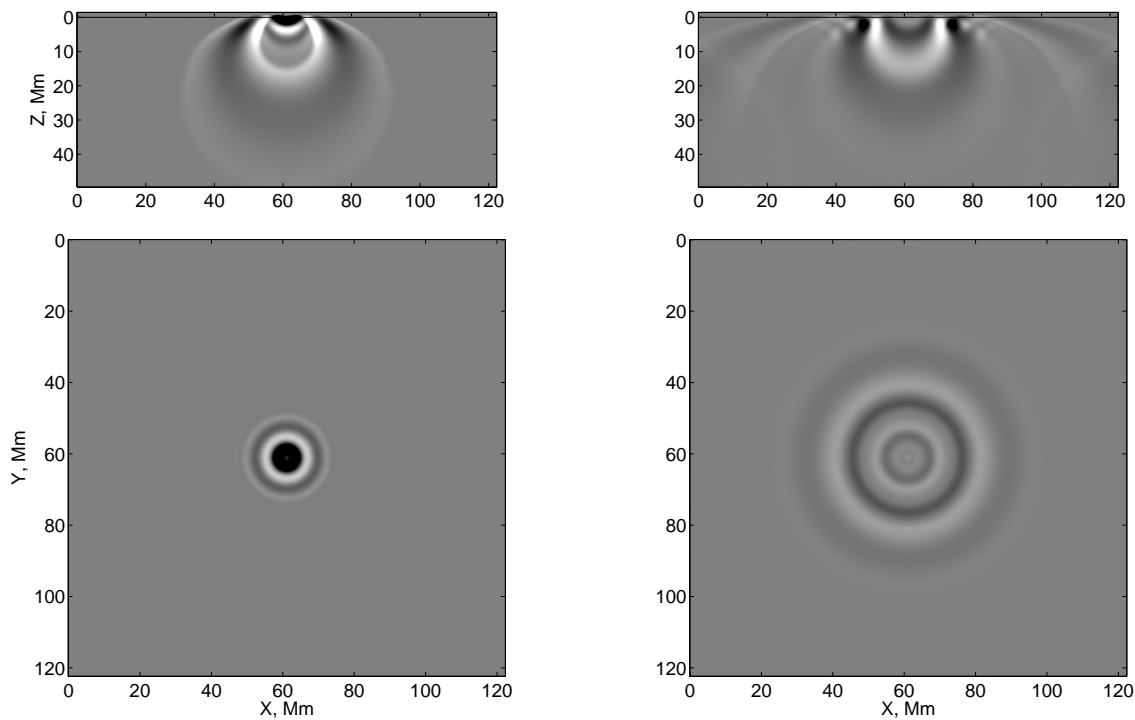} \caption{Snapshots of the density
perturbation from spherically symmetric Gaussian pulse of a
z-component of force at $t=11.7$ min (left column) and $t=21.7$ min
(right column). The top row represents the vertical slices of the
computational domain, the bottom row shows the horizontal slices at
a height of 350 km above the photosphere. Thin horizontal line at
$z=0$ represents the photosphere. The depth of the source equals 3.5
Mm below the photosphere. \label{SS_3D_fig}}
\end{figure}

Numerical simulations of propagation of waves in 3D from a single
source inside the Sun are shown in Figure~\ref{SS_3D_fig}. The
Brunt-V\"{a}is\"{a}l\"{a} frequency of the standard solar model with
smoothly joined chromosphere was modified near the surface to make
the model stable against convection. Such modified model was chosen
as the background model. Non-reflecting boundary conditions (PMLs)
were established at the top and bottom boundaries. The top layer was
established at the height of 500 km above the photosphere in the
region of the temperature minimum. This layer absorbs all waves with
frequencies higher than the acoustic cut-off frequency which pass to
the chromosphere and do not affect reflection of waves with lower
frequencies, because these waves are reflected from layers below the
photosphere. Lateral boundary conditions are periodic. The
computational domain of size 120$\times$120$\times$50 Mm${}^3$ was
covered by the uniform grid of size 720$\times$720$\times$300 nodes
with spatial intervals $\Delta x=\Delta y=\Delta z = 170$ km. The
time step $\Delta t=1$ sec was chosen from stability condition. The
Gaussian spherically symmetric pulse source of z-component of force
\begin{equation}\label{SRC3D_fun}
\Phi(x,t)=\exp\left[-
\left(\frac{\mi{r-r}_{src}}{\sigma}\right)^2\right]\sin(\omega_0
t),\quad 0\leq t\leq 2\pi/\omega_0.
\end{equation}
with $\sigma=0.4$ Mm was placed at the depth of 3.4 Mm below the
photosphere. For such a choice of $\sigma$ semi-width of the source
equals approximately 4 grid nodes. The time dependent part is just
one period of sin-function with $\omega_0=2.5$ mHz.
Figure~\ref{SS_3D_fig} shows snapshots of the density perturbation
from such a source at $t=11.7$ min (left column) and $t=21.7$ min
(right column). The top row represents the vertical slices of the
computational domain, the bottom row shows the horizontal slices at
a height of 350 km above the photosphere. The thin horizontal line
at $z=0$ represents the photosphere. The left column shows the
disturbance from the direct wave, generated by the source. The right
column shows the wave reflected from the photosphere. The reflected
wave front is broader and has less amplitude, because our source has
finite lifetime (one period of sine) and generates high frequency
waves which pass through the photosphere. Such waves are absorbed by
the top non-reflecting layer and do not make a contribution to the
amplitude of the reflected wave.

\section{Results and Discussion}
\begin{figure}
\epsscale{1.0} \plotone{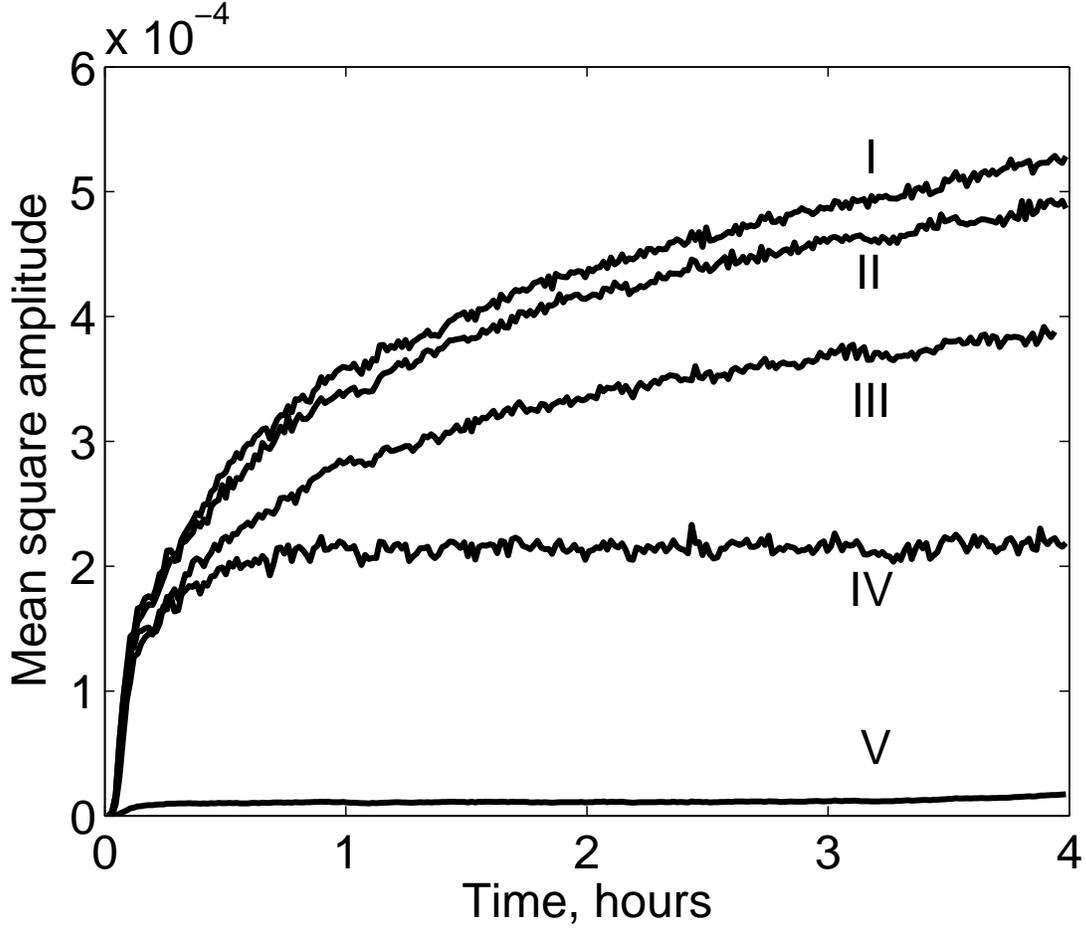} \caption{Wave amplitude averaged
along the horizontal plane at the height of 300 km above the
photosphere for different heights of the top boundary and different
damping coefficients. The curve I corresponds to the high top
boundary, established at 1750 km above the photosphere without any
additional damping. The curves II, III, V correspond to the same
boundary conditions, but different values of damping coefficient
$\sigma_d$ = 0.3, 0.6, 1.0. The curve IV corresponds to the low top
boundary, established at 500 km without artificial damping.
\label{MeanAmpl_fig}}
\end{figure}

\begin{figure}
\epsscale{1.0} \plotone{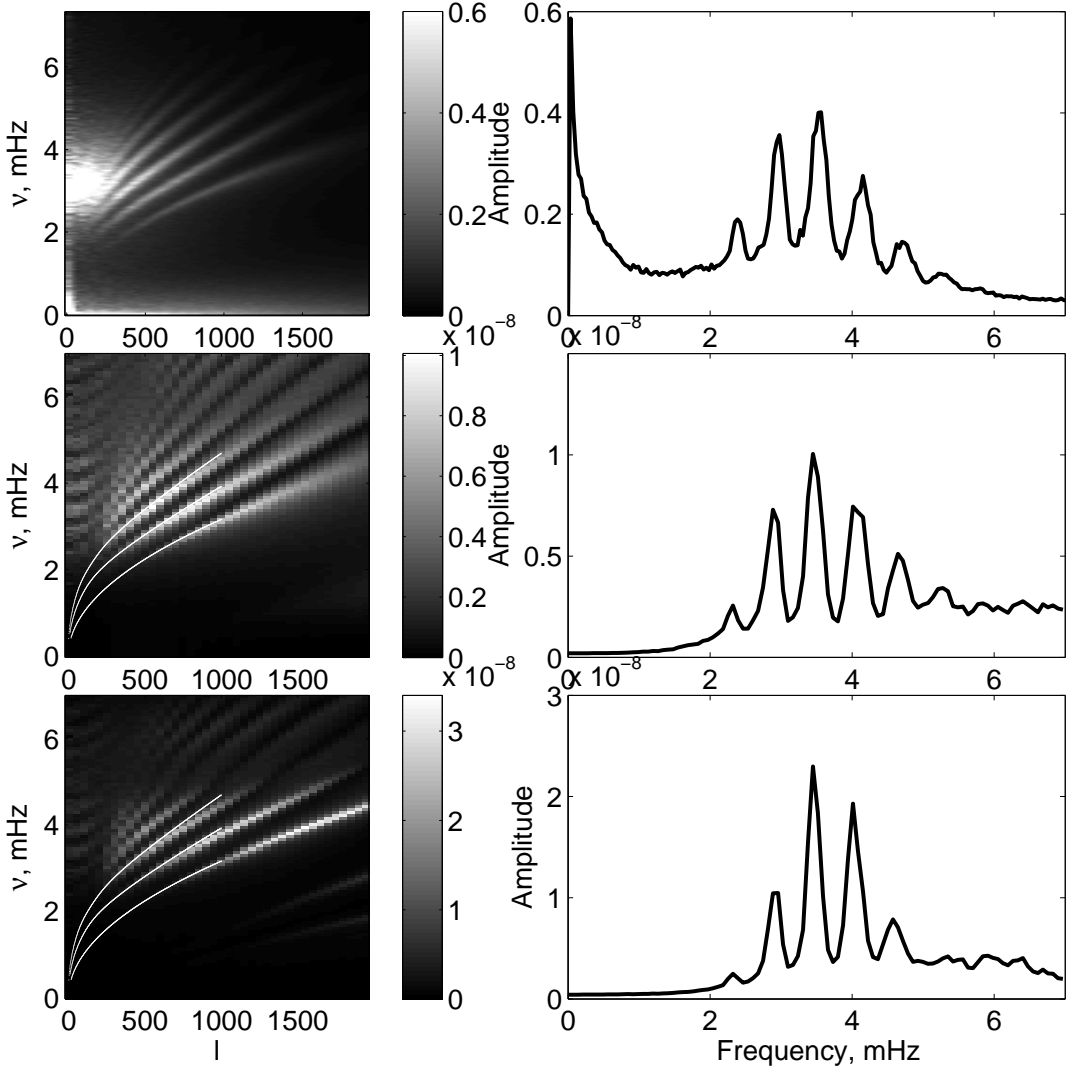} \caption{Acoustic spectra obtained
from observations and simulations with different heights of the top
boundary (the left panes). Starting from the top: observations,
simulations with $h_{top}$=500 km, simulations with $h_{top}$=1750
km. The thin white curves on the left panes show position of
observational ridges for $f$, $p_1$, and $p_2$ modes. The right
panes show cuts of k-$\omega$ diagrams from left panes at l=584. For
simulations with high top boundary without additional damping
(bottom row) acoustic modes trapped in the domain distort the shape
of the acoustic spectrum. \label{kw_fig}}
\end{figure}

The developed numerical method and code have been used for
simulation of the acoustic wave field generated by multiple acoustic
sources inside the Sun. We found that the height of the PML affects
absorbing properties of the top boundary and the shape of the
acoustic spectrum (k-$\omega$ diagram). Reflection from the top
boundary is a wave process. The waves are reflected not from the
fixed level but from some vertical region. Region with the acoustic
cut-off frequency greater than the wave frequency $\omega_c >
\omega$ acts as a potential barrier for such waves. Even if the wave
frequency is less than the acoustic cut-off frequency waves
penetrate to this region with exponentially decaying amplitude. If
the thickness of the barrier is finite waves can leak through it.
This process is similar to the tunneling effect in quantum
mechanics. This happens in the real Sun as well. We studied behavior
of the solution for different heights of the top boundary. All
depths and heights are calculated from the level of the photosphere
($r=R_{sun}$ in the Christensen-Dalsgaard's standard solar model S).
The background model varies fast in the region above the temperature
minimum. To be able to simulate propagation of acoustic waves in the
chromosphere we were forced to reduce the vertical spatial step to
$\Delta z =50$ km to preserve the numerical stability. To keep the
horizontal size of the domain as in previous simulations without
significant increasing number of grid nodes the horizontal spatial
steps were chosen 3 times bigger $\Delta x = \Delta y = 3\Delta z =
150$ km. To satisfy the Courant stability condition for the explicit
scheme, the time step was reduced to $\Delta t=0.68$ sec. The
computational domain of size 122.2 Mm $\times$ 122.2 Mm $\times$ 32
Mm is covered by the uniform grid of size 816$\times$816$\times$640.
Sources of z-component of force with random frequencies were
randomly distributed at the depth of 350 km. Sources are initiated
at random moments of time (one source per time step) and depend on
time as Ricker's wavelet with central frequency from range 2$\div$8
mHz. In the Sun we have some damping due to the turbulent viscosity.
We simulated this additional viscosity by adding a damping term
$-\sigma_d q_z$ to the right hand side of z-momentum equation in the
region above the photosphere and smoothly fading to zero below. The
time dependence of RMS (root mean square) wave amplitude averaged
along the horizontal plane at the height of 300 km above the
photosphere for different heights $h_{top}$ of the top boundary and
different values of the damping coefficient $\sigma_d$ is shown in
Figure~\ref{MeanAmpl_fig}. The RMS amplitude for the high PML
established at the height of $h_{top}=1750$ km without additional
damping $\sigma_d$ = 0 is shown by the solid curve I. The RMS
amplitudes for the same height of the top PML and $\sigma_d$ = 0.3,
0.6, and 1.0 are plotted by solid lines II, III, and V
correspondingly. In the last case the RMS amplitude reaches an
equilibrium state. The curve IV corresponds to the low PML,
established at the height of 500 km above the photosphere without
additional damping in the inner region $\sigma_d$ = 0. RMS amplitude
reaches an equilibrium state in this case as well, because the
acoustic modes leak through the acoustic potential barrier and their
exponential tails reach the top absorbing boundary, which adds an
additional damping and stabilizes the amplitude. The top boundary
with the height of 1750 km is set high enough and does not affect
modes with frequencies less than the acoustic cut-off frequency.
Energy is continuously pumped to the system by acoustic sources. The
bottom boundary is deep enough that some modes resolved by numerical
scheme have turning points above the bottom boundary. Such modes are
trapped in the domain, the total energy increases, and the RMS
amplitude does not reach an equilibrium state. This distorts the
acoustic power spectrum and changes the amplitude ratio of trapped
modes and modes that can be absorbed at the top and/or bottom
boundaries. The left panes in Figure~\ref{kw_fig} show the acoustic
power spectra obtained from observations (top), simulations without
damping with low (500 km) top boundary (middle), and high (1750 km)
top boundary (bottom). The right panes show the vertical cuts of
corresponding k-$\omega$ diagrams at $l=584$. The  bottom left pane
(high top boundary) shows presence of g- modes in simulations. They
appear because our background model is convectively stable in the
thin layer below the photosphere. In the real Sun, this layer is
convectively unstable, and thus the g- modes do not propagate.
Energy leakage through the acoustic potential barrier in the case of
low top boundary (middle row) damps all modes uniformly and does not
change the shape of the acoustic spectrum. So the height of the top
boundary can be used for controlling the damping rate without
distortion of the acoustic spectrum.

\section{Conclusion}
Developed linear 3D code for propagation of acoustic waves inside
the Sun uses the realistic equation of state and realistic
non-reflecting boundary conditions which permits to simulate
accurately reflection of waves from the top boundary. Waves with
frequencies less then the acoustic cut-off frequency are reflected
from the photosphere, and waves with higher frequencies pass to the
chromosphere. The top non-reflecting boundary absorbs such waves.
Establishing the top boundary high enough in the chromosphere leads
to the trapping of some modes in the domain and increasing their
amplitudes with time, which distorts the shape of the acoustic
spectrum. Energy leakage through the acoustic potential barrier in
case of the low (500 km) top boundary leads to an additional uniform
damping of all modes which stabilizes the amplitude and does not
distort the spectrum. The height of the non-reflecting top boundary
can be used as a parameter for controlling of the damping rate in
the system. The acoustic spectrum obtained from simulated wave field
shows existence of p-, and f-modes.  The simulated acoustic spectrum
is good agreement with observations.

This code has been used by \citep{Parchevsky2006} to model the
effects of non-uniform spatial distribution of acoustic sources in
sunspot regions. Their results showed that this effect can explain
at least a half of the observed amplitude reduction in sunspots.
 The code can be used for studying details of
interaction of waves with inhomogeneities of solar structure and for
producing artificial data for testing an accuracy of helioseismic
inversion as well, as for studying of propagation of acoustic waves
in the chromosphere and reflecting properties of the photosphere.
Future simulations will include subsurface flows and magnetic field.
\section{Acknowledgements}
This research is supported by the Living With the Star NASA grant
NNG05GM85G. The calculations were performed on Columbia
supercomputer at NASA Ames Research Center (NASA Advanced
Supercomputing Division).

\end{document}